\documentclass[aps,prl,twocolumn,floatfix,showpacs,showkeys,amssymb]{revtex4}

\usepackage{graphicx}

\usepackage{amsmath}

\usepackage{amsfonts}

\usepackage{amssymb}

\begin{document}
\topmargin -15mm

\def\ba{\begin{array}}

\def\ea{\end{array}}

\def\be{\begin{equation}\begin{array}{l}}

\def\ee{\end{array}\end{equation}}

\def\bea{\begin{equation}\begin{array}{l}}

\def\eea{\end{array}\end{equation}}

\def\f#1#2{\frac{\displaystyle #1}{\displaystyle #2}}

\def\om{\omega}

\def\Om{\Omega}

\def\omm{\omega^a_b}

\def\we{\wedge}

\def\de{\delta}

\def\De{\Delta}

\def\va{\varepsilon}

\def\omb{\bar{\omega}}

\def\la{\lambda}

\def\vv{\f{V}{\la^d}}

\def\si{\sigma}

\def\t{T_+}

\def\v{v_{cl}}

\def\m{m_{cl}}

\def\n{N_{cl}}

\def\bi{\bibitem}

\def\c{\cite}

\def\sa{\sigma_{\alpha}}

\def\ua{\uparrow}

\def\da{\downarrow}

\def\mua{\mu_{\alpha}}

\def\ga{\gamma_{\alpha}}

\def\g{\gamma}

\def\ora{\overrightarrow}

\def\pa{\partial}

\def\ov{\ora{v}}

\def\al{\alpha}

\def\bt{\beta}

\def\R{R_{\rm eff}}

\def\muu{\f{\mu}{ed}}

\def\E{\f{edE(\tau)}{\om}}

\def\t{\tau}

\title{Electronic states in a magnetic quantum-dot molecule: phase
transitions and spontaneous symmetry breaking} 

\author{Wei Zhang, Tianming Dong, Alexander O. Govorov}

\affiliation{Department of Physics and Astronomy,
Ohio University, Athens, OH 45701-2979 }
\date{\today } 

\begin{abstract}
We show that a double quantum-dot system made of diluted magnetic
semiconductor behaves unlike usual molecules. 
In a semiconductor
double quantum dot or in a diatomic molecule, the ground state of
a single carrier is described by a symmetric orbital. In a
magnetic material molecule, new ground states with broken symmetry
can appear due the competition between the tunnelling and magnetic
polaron energy. With decreasing temperature, the ground state
changes from the normal symmetric state to a state with
spontaneously broken symmetry. Interestingly, the symmetry of a
magnetic molecule is recovered at very low temperatures. A
magnetic double quantum dot with broken-symmetry phases can be
used a voltage-controlled nanoscale memory cell.
\end{abstract}



\maketitle




In diluted  magnetic semiconductors,  
the spins of static impurities 
interact through mobile carriers and can form 
a ferromagnetic state \cite{Furdyna,FerroBulkZener}. Since the
carrier density in semiconductors is a voltage-tunable parameter,
the ferromagnetic state also becomes controlled by the voltage
\cite{FerroBulk}. For technology, this may give an important
advantage compared to the convectional memories based on
ferromagnetic metals and controlled by the magnetic field. When a
single confined carrier interacts with magnetic impurities, it
forms a new stable state, called a magnetic polaron \cite{wolf}.
One important class of confined nanostructures is the quantum dots
(QDs) where the number of trapped carries can be easily changed by
the voltage applied to the top gate \cite{capacity}. Information
in a single QD made of magnetic semiconductor can be stored in the
form of spin polarization and therefore such a QD can be viewed as
a nanoscale magnetic memory element. Magnetic QDs as memory
elements have some properties, that look very attractive for the
technology: (1) small sizes, (2) small number of carriers, and (3)
voltage control of the number of carriers. Currently, the physics
of electronic states in magnetic QDs is an active field of
research
\cite{Kulak-Leigh-Molenk,theory,theory1,theory2,govorovPRB2005}.
Here we make a logical step from a single magnetic nanocrystals
toward QD molecules and show that a magnetic QD pair has unique
physical properties that may also be  useful for device
applications.

\begin{figure}[tbh]
\includegraphics*[width=0.8\linewidth, angle=90]{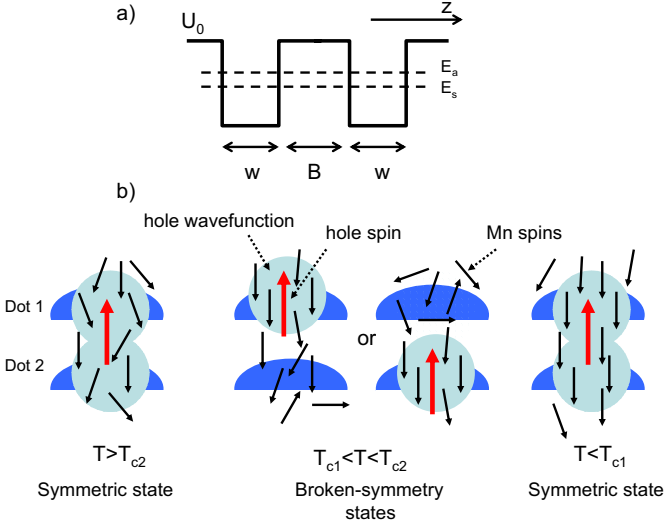}
\caption{(a) The potential $u(z)$. (b) Schematics of the magnetic
QD molecule and three magnetic-polaron phases with a single hole.}
\label{fig1}
\end{figure}

In this letter, we consider a magnetic double QD with one hole.
Using the mean field theory, we calculate the physical properties
of magnetic polarons formed due to the Mn-hole exchange
interaction. We find that this system undergoes two phase
transitions (Fig.~\ref{fig1}). At high temperatures, the kinetic
tunnelling energy of the hole is larger than the magnetic polaron binding energy
 and a symmetric state (with equal hole probabilities
for both dots) is realized. With decreasing temperature ($T$), the
local magnetic energy becomes large enough to trap the hole in one
of the dots. This self-trapped process spontaneously breaks the
symmetry of the system. At very low temperature, the local Mn-spin
polarization can become so strong that the symmetry is recovered.
Our broken-symmetry polaron has several unique properties: it appears
in the confined geometry, is voltage-controllable, and may vanish
at very low $T$.
The above phase transitions occur also with changing the energy
barrier between the dots. Since a double QD can be controlled by
the gate voltage, the above phases can be prepared and read using
electrical means.

{\bf Model}. The Hamiltonian of the system composed of two
magnetic QDs and one hole has a form

\be H_{hh}=\f{\bf \hat{P}^2}{2m_{hh}}+U_0({\bf
R})-\f{\beta}{3}\hat{j}_z\cdot \hat{S}_z, \ee where ${\bf R}=({\bf
r}, z)$, ${\bf r}$ and $z$ are the in-plane vector and the
vertical coordinate, respectively; $\hat{j}_z$ is the z-component
of the hole spin and $j_z=\pm 3/2$; the operator $\hat{S}_z=\sum_i
\hat{ S_{i,z}\de(\bf R-R_i)}$ is the spin operator associated with
the Mn subsystem; here $\hat{S_{i,z}}$ and ${\bf R}_i$ are the
single-impurity spin and position, respectively; $i$ is the
Mn-impurity index. In this study, we will model the in-plane hole
motion by a parabolic potential and the vertical motion with a
square double well \cite{PawelScience}, i.e. $U_0({\bf
R})=u(z)+m_{hh}\om_0^2r^2/2$, where $u(z)$ is a double well
potential in the $z$-direction (Fig.~\ref{fig1}a) and $\om_0$ is
the in-plane frequency, $m_{hh}=m^*$ is the mass of heavy holes.
In our model describing disk-like self-assembled QDs, the
light-hole states are assumed to be split from the lowest
heavy-hole states. In the absence of Mn impurities, the hole
wavefunction can be written as
$\psi_J=\phi_j(z)\varphi_{n_x,n_y}(r)$, where $J=(j,n_x,n_y)$,
$n_{x(y)}=0,1,2...$, and $j=(s,a,k_z)$ label the discrete ($s$ and
$a$) and continuous states ($k_z$) related to the $z$-motion.
Above we assumed that our QD pair has only two bound states,
symmetric (s) and antisymmetric (a).

Within the mean field theory, the Hamiltonian becomes

\be \bar {H}_{hh}=\f{\bf \hat{P}^2}{2m^*}+U_0({\bf
R})-\f{\beta}{3}x_{Mn}N_0\hat{j}_z\cdot \bar{S}_z, \ee where
$x_{Mn}$ is the reduced Mn concentration, $N_0$ is the number of
cations per unit volume, and $\bar{S}_z({\bf R})$ is the local
average Mn spin

\be \bar{S}_z({\bf R})=S B_S [\f{\beta/3 \bar{j}_z ({\bf
R})S}{k_B(T+T_0)}], \label{S} \ee where S=5/2, $B_S$ is the
Brillouin function, $T_0$ describes the effect of
anti-ferromagnetic interaction between Mn impurities, $\bar
{j}_z({\bf R})=<\psi({\bf R'})|\hat{j}_z \de({\bf R-R'})|\psi (\bf
R')>$, where $\psi$ is the hole wave function.

In the spirit of mean field theory, the hole moves in the presence
of the effective spin-dependent  potential,

\be \bar{U}=U_0+\delta U, \ \delta U({\bf
R})=-\f{\beta}{3}x_{Mn}({\bf R})N_0\hat{j}_z\cdot \bar{S}_z.
\label{U}\ee

The corresponding ground state wave function has the form:
$\psi=\psi_G |\uparrow>$, where $|\uparrow>$ is the hole state
with $j_z=+3/2$. The spatial wave function $\psi_G$ should be now
determined. To the lowest order perturbation theory (PT)
\cite{notePT}, the ground-state wave function can be written as
$\psi_G^0=\phi_G(z)\varphi_{0,0}(r)$, with

\be \phi_G=a \phi_s +\sqrt{1-a^2}\phi_a, \ee where $\phi_s$ and
$\phi_a$ are the bound states in the double well. Then, the real
parameter $a$ should be determined by the variational method. From
eqs.~\ref{S}, \ref{U}, and $\bar {j}_z({\bf
R})=\f{3}{2}\psi_G^2({\bf R})$ we can obtain the polaron binding
energy and the total energy

\be E_b(T)=-\f{\beta}{2}x_{Mn} N_0 S
\int d^3R[(\psi_G^0)^2 B_S(\f{\beta/3 \bar{j}_z ({\bf R}) S}{k_B(T+T_0)})], \\
E_{tot}^1(a)=a^2 E_s+(1-a^2)E_a+E_b. \label{Eb} \ee The general
variational wave function can be written as
$\psi_G=\phi_G(z)\varphi_{0,0} ({\bf r})+\sum_J a_J \psi_J$, where
the second term will be considered as perturbation. The zero-order
degenerate PT provides us with two orthogonal wavefunctions
$\psi_G^0$ and $\psi_E^0$ \cite{notePT}. The corresponding
first-order PT contribution to the energy is given by
eq.~\ref{Eb}. Then, the second-order correction to the energy
becomes

\be \de E=\sum_{\Gamma}\f{|\delta
U_{G,\Gamma}(\psi_G^0)+\f{1}{2}(\f{\pa \delta U(\psi_G^0)}{\pa
a_{\Gamma}})_{G,G}|^2}{E_{tot}^1-E_{\Gamma}}, \label{dE}\ee where
$\Gamma=\{\gamma,n_x,n_y\}$, $\gamma=(G,E,k_z)$, $n_x,
n_y=\{0,1,2,...\}$, $\Gamma\not =\{G,0,0\},\{E,0,0\}$; here
$E_G^1$ and $\psi_G^0$ describe the ground state calculated within
the lowest-order PT and $k_z$ labels the delocalized states  for
the $z$-motion.

\begin{figure}[tbh]
\includegraphics*[width=0.7\linewidth, angle=90]{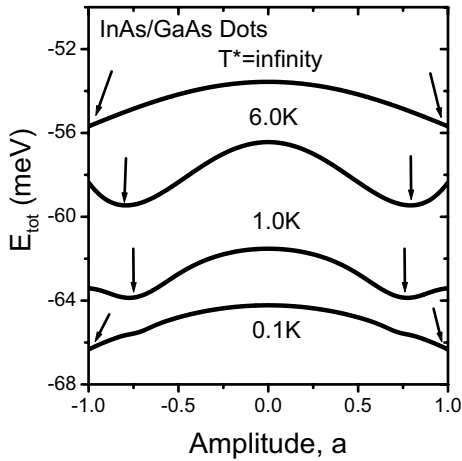}
\caption{The total energy of magnetic polaron as a function of the
wavefunction parameter $a$ at different $T^*$. The arrows show the
ground states of the system.} \label{fig2}
\end{figure}

{\bf Phase transition as a function of temperature}. To find the
variational ground state, we calculate the total energy
$E_{tot}(a)=E_{tot}^1(a)$ as a function of $a$ at various
temperatures (Fig.~\ref{fig2}). The quantity $a$ is the hole
amplitude related to the symmetric state.  The material parameters
are the following: $\beta N_0=-1.8~eV$, $N_0=23~nm^{-3},
l_0=\sqrt{\hbar/m^*\omega_0}=2.5~nm$, $b=4~nm$, $w=3~nm$,
$U_0=90~meV$, $m=0.38~m_0$, $x_{Mn}=0.005$, and $R_{Mn}=15.0~nm$.
This parameter set represents a InAs/GaAs QD doped with Mn
impurities. One can clearly see three different phases. For high
$T$, the ground state corresponds to $a^2=1$, i.e., the ground
state is symmetric. With decreasing $T$, the ground changes from
$a^2=1$ to almost $a^2=0.5$. It is a state of hole staying mostly
in one dot. Thus, the symmetry of the system becomes broken. For
very low $T$, the symmetry is recovered.

\begin{figure}[tbh]
\includegraphics*[width=0.9\linewidth]{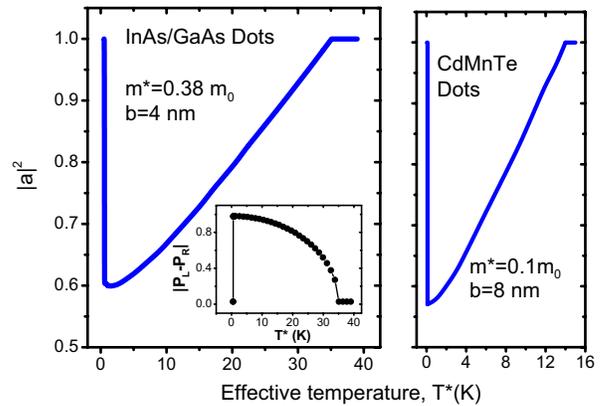}
\caption{ The wavefunction factor $a^2$ as a function
of the effective temperature; the function $a^2(T^*)$ shows the
appearance of broken-symmetry phase ($a^2<1$). Inset: The
effective order parameter (degree of asymmetry) as a function of
$T^*$.} \label{fig3}
\end{figure}

The probability $a^2$ (Fig.~\ref{fig3}) as a function of the
effective temperature $T^*=T+T_0$ clearly shows the existence of
three phases. We may consider the difference of hole probabilities
of being inside the dots
$\delta=|P_{upper}-P_{lower}|=2a\sqrt{(1-a^2)}$ as an order
parameter (as shown in the inset of Fig.~\ref{fig3}). The order
parameter $\delta$ is nonzero for the broken-symmetry phase at
intermediate $T^*$ and is zero for the symmetric state at high and
low $T^*$.

Basically, the physical picture is the following: the interaction
between the hole and Mn impurities leads to the ferromagnetic
ordering of Mn-spins inside the double QD. The resulting polaron
state tends to be as localized as possible because the magnetic
binding energy increases with decreasing the localization strength
of hole. The latter can be seen from eq.~\ref{Eb}: the Brillouin
function strongly increases with increasing the hole wavefunction
amplitude, $\psi^2$. Therefore, in our system there is a
competition between the kinetic energy (tunnel splitting) and the
magnetic energy. At high $T$, due to the thermal fluctuations, the
interaction between the hole and Mn impurities is weak and the
kinetic energy wins. In this case, a symmetric ground state is
realized. With decreasing $T$, the Mn-hole interaction becomes
stronger and, below some critical temperature $T_{c2}$, the
magnetic localization effect overcomes the tunnelling. The hole is
now trapped in one of the dots since such a spatial configuration
lowers the total energy. However, for very low $T^*$, the hole is
able to polarize the Mn spins inside the QD pair much stronger and
the magnetic potential $\delta U$ becomes almost flat. Therefore,
the symmetric orbital state is recovered. There is an obvious
condition for the existence of the low-T phase: $T_{c1}^*>T_0 (T_{c1}>0)$.
In other words, for the existence of the low-T phase, the Mn-hole
interaction must be strong enough to overcome the
antiferromagnetic interaction. Fig.~\ref{fig1}b shows
schematically the three phases.

We should point out that the symmetric polaron at low $T$ has a
much stronger Mn spin polarization than the symmetric state at
high $T$. The temperature $T^*_{c2}$ corresponds to a smooth,
type-II phase transition, while the low-T transition is sharp and
can be classified as a type-I phase transition. The low-T
transition occurs as a jump between the broken-symmetry phase
($a\approx\pm1/\sqrt{2}$) and the symmetric state ($|a|=1$).

\begin{figure}[tbh]
\includegraphics*[width=0.9\linewidth]{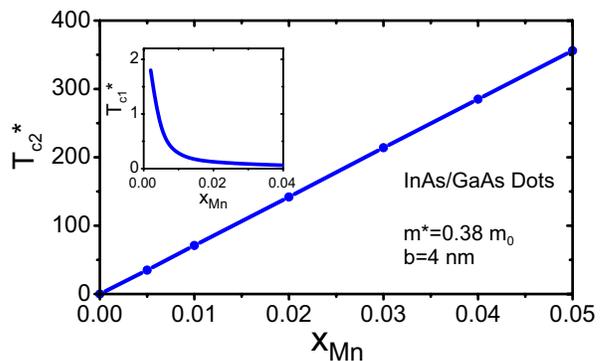}
\caption{Temperatures $T^*_{c2}$ and $T^*_{c1}$ as functions of the Mn
concentration. } \label{fig4}
\end{figure}

The temperature $T^*_{c2}$ can be found analytically by expanding
the total energy in the vicinity of the type-II phase transition:

\be k_B T^*_{c2}=\f{\beta^2 N_0 S(S+1)}{6(E_a-E_s)}\int d^3R~ x_{Mn}
(3\phi_s^2 \phi_a^2-\phi_s^4). \label{Tc2} \ee Fig.~\ref{fig4}
shows numerically calculated $T_{c2}^*(x_{Mn})$ (dots) and the
data obtained from eq.~\ref{Tc2} (line). Our numerical and
analytical results agree well. The temperature $T^*_{c2}\propto
x_{Mn}$. This linear dependence may be understood qualitatively.
For a higher Mn-consecration, stronger thermal fluctuations
(higher $T$) are needed to destroy the broken-symmetry polaron
phase. In contrast to $T^*_{c2}$, the temperature $T^*_{c1}$ decreases
with $x_{Mn}$ (approximately as $1/x_{Mn}$)(Fig.~\ref{fig4}).
Qualitatively, such a dependence can be explained as follows. The
low-T symmetric phase appears when the Mn spins are strongly
polarized. Therefore, a lower $T^*_{c1}$ is needed to polarize a
larger number of Mn spins.

As discussed above, the transition temperature $T^*_{c2}$ depends on
the competition between the kinetic and magnetic energies and
qualitatively corresponds to the condition $|E_a-E_s|\sim \delta
U$. $T^*_{c2}$ depends on a material. In Fig.~\ref{fig3} we show the
results for a CdMnTe QD. For the parameters, we use: $\beta
N_0=-1.0~eV$, $N_0=15~nm^{-3}$, $l_0=4~nm$, $b=8~nm$, $w=5~nm$,
$U_0=150~meV$, $m=0.1~m_0$, $x_{Mn}=0.005$, and $R_{Mn}=15.0~nm$.
The effective transition temperature is $14~K$ and larger than the
typical antiferromagnetic CdMnTe temperature $T_0=3.6~K$.

\begin{figure}[tbh]
\includegraphics*[width=0.5\linewidth, angle=90]{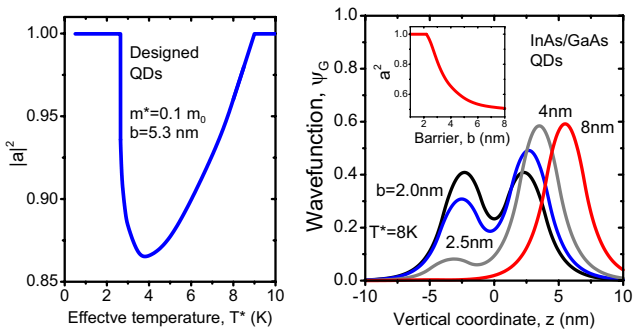}
\caption{Left: Phase transitions in a specially designed magnetic
QD pair with a larger $T^*_{c1}$. Right: The ground-state wave
function for different barrier widths for the Mn-doped InAs/GaAs
QD pair. Inset: phase transition as a function of the barrier
width. } \label{fig5}
\end{figure}

One may notice that the transition temperature $T^*_{c1}$ can be
very small (Fig.~\ref{fig4}, inset). It may be impossible to
observe the third phase ($T^*_{c1}<T_0$), if some anti-ferromagnetic
interactions are present. However, due to to the tunability of QD
systems, we may design a system with a relatively high $T^*_{c1}$.
Fig.~\ref{fig5} shows the phase transitions in a designed system
with $T^*_{c1}\sim 2~K$. For the parameters, we took: $\beta
N_0=-2.2~eV$, $N_0=10~nm^{-3}$, $l_0=4~nm$, $B=5.3~nm$, $w=5~nm$,
$U_0=150meV$, $m=0.1~m_0$, $x_{Mn}=0.005$, and $R_{Mn}=3.0nm$.

In recent experiments \cite{TwoDots}, self-assembled double QDs
were built into transistor structures with top and back contacts.
By using applied voltage in these structures one can change the
number of trapped carriers and also modify their wave functions.
Two stable magnetic states in our system (Fig.~\ref{fig1}b) can be
considered as a bit of classical information. In these states, the
hole mostly resides either in the upper dot or in the lower dot.
Using the gate voltage, our system can be prepared in the desired
state. The following electrical readout of the polaron state can
be done, for example, using capacitance spectroscopy
\cite{capacity}.

{\bf Quantum phase transition}. We may look at the physics from
another point of view. The kinetic energy of the hole, i.e. the
hopping between the dots, is related to the width of the barrier
between QDs and thus can be tuned by appropriate growth process.
In Fig.~\ref{fig5}, we show the ground-state wavefunctions for
various widths of the barrier. With increasing of the width, the
ground states become more asymmetric. It is clear that there is a
quantum phase transition with the increasing of barrier width. The
critical width is around $2~nm$ (see inset of Fig.~\ref{fig5}).

{\bf Second order corrections to energy}. All the previous discussions are
based on the perturbation theory applied to the nonlinear
Schr\"odinger equation. The second-order correction to the energy
can be calculated by eq.~\ref{dE}. For the sets of parameters used
above, it is of a few percent of the first-order result. We also
found numerically that, in the vicinity of transition temperatures
$T^*_{c1}$ and $T^*_{c2}$, the second-order corrections are even
smaller \cite{note}. This allows us to give the results for
$T^*_{c1}$ and $T^*_{c2}$ in a wider range of $x_{Mn}$ (see
Fig.~\ref{fig4}).


In conclusion, we have studied a magnetic QD molecule with a
single hole. With decreasing temperature, the magnetic polaron
ground state undergoes two phase transitions. At a higher
temperature, the normal symmetric state turns into the polaron
state with broken symmetry. At lower temperatures, a symmetric
polaron state can be recovered.  Our results suggest that a
magnetic QD molecule can be used as a nanoscale magnetic memory
cell controlled electrically.

{\bf Acknowledgments}
This work was supported by the BNNT Initiative at Ohio University.


\end{document}